\documentclass{ws-procs10x7}
\usepackage{balance}

\newcolumntype{d}[1]{D{.}{.}{#1}}

\def\Journal#1#2#3#4{{\it #1} {\bf #2}, #3 (#4)}

\newcommand{\3}{\mbox{${\bf \underline{3}}$}}
\newcommand{\s}{\mbox{${\bf \underline{1}}$}}
\newcommand{\spr}{\mbox{${\bf \underline{1}'}$}}
\newcommand{\sppr}{\mbox{${\bf {\underline{1}''}}$}}

\makeindex
\begin{document}

\title{$A_4$ SYMMETRY BREAKING SCHEME FOR UNDERSTANDING\\
QUARK AND LEPTON MIXING ANGLES}

\author{R. R. VOLKAS$^*$}

\address{School of Physics, Research Centre for High Energy Physics, The 
University of Melbourne, Victoria 3010, Australia\\$^*$E-mail: r.volkas@physics.unimelb.edu.au}


\twocolumn[\maketitle\abstract{The neutrino mixing matrix has been measured to be of a form
consistent with tribimaximal mixing, while the quark mixing matrix is almost diagonal.  A scheme
based on flavour $A_4$ symmetry for understanding these patterns simultaneously is presented.}
]

\section{Tribimaximal mixing}

The current neutrino oscillation data are well described by the
following MNSP mixing matrix:
\begin{equation}
\left(\begin{array}{ccc}
\frac{2}{\sqrt{6}} & \frac{1}{\sqrt{3}} & 0 \\
-\frac{1}{\sqrt{6}} & \frac{1}{\sqrt{3}} & -\frac{1}{\sqrt{2}} \\
-\frac{1}{\sqrt{6}} & \frac{1}{\sqrt{3}} & \frac{1}{\sqrt{2}}
\end{array}\right)
\end{equation}
This is called ``tribimaximal mixing''\cite{tribi}. Complex phases can also be introduced.
It is significant that the entries are square roots of fractions formed from
{\it small} integers\cite{zee-small-integer}, and it is suggestive of a flavour symmetry.
It also motivates that the flavour structure required to understand mixing should be divorced from
whatever physics is needed to understand the mass eigenvalues, because the latter do not at this stage seem
to show suggestive patterns.

We shall call a matrix ``form diagonalisable (FD)'' if its (left) diagonalisation
matrix is formed from definite numbers while its eigenvalues are 
free parameters.\cite{fd} A simple $2 \times 2$ example is
\begin{equation}
\left(\begin{array}{cc}
m_1 & m_2 \\ m_2 & m_1 \end{array}\right)
\end{equation}
whose diagonalisation matrix gives two-fold maximal mixing, while its eigenvalues are
arbitrary and depend on $m_{1,2}$.  This matrix has a $Z_2$ structure, and arose
in the mirror matter model.\cite{mm} A relevant $3 \times 3$ example is
\begin{equation}
\left( \begin{array}{ccc}
m_1 & \ \ m_2 & \ \ m_3 \\
m_1 & \ \ \omega\, m_2 & \ \ \omega^2\, m_3 \\
m_1 & \ \ \omega^2\, m_2 & \ \ \omega\, m_3
\end{array} \right)
\end{equation}
where $\omega \equiv e^{i2\pi/3}$ is a cube root of unity. It is equal to
\begin{equation}
U(\omega) \left( \begin{array}{ccc} \sqrt{3}m_1 & 0
& 0 \\ 0 & \sqrt{3} m_2 & 0 \\ 0 & 0 & \sqrt{3}m_3
\end{array} \right)
\end{equation}
where the left-diagonalisation matrix is ``trimaximal'':
\begin{equation}
U(\omega) = \frac{1}{\sqrt{3}} \left( \begin{array}{ccc} 1 & 1 & 1
\\ 1 & \omega & \omega^2 \\ 1 & \omega^2 & \omega
\end{array} \right).
\end{equation}
Since the MNSP matrix $V_{MNSP} = V^{e\dagger}_L\, V^{\nu}_L$ is the product of
two diagonalisation matrices, we observe that tribimaximal mixing is obtained
from $U(\omega)^{\dagger} V_L^{\nu}$ when
\begin{equation}
V_L^{\nu} = \frac{1}{\sqrt{2}} \left( \begin{array}{ccc} 1 & 0 &
-1 \\ 0 & \sqrt{2} & 0 \\ 1 & 0 & 1
\end{array} \right).
\end{equation}
which is the previous $Z_2$ structure in the $(1,3)$ subspace.

\section{$A_4$ scheme and tree-level results}

$A_4$ is the set of even permutations of four objects.\cite{a4,af}  It has 12 elements:
$1, c, a=c^{-1}, r_{1,2,3}=r_{1,2,3}^{-1}, r_i c r_i, r_i a r_i$, where
$\{1,c,a\}$ form $C_3 = Z_3$ subgroup, $\{1,r_i\}$ form $Z_2$ subgroups 
(see\cite{zee-small-integer} for notation).
Its irreducible reps are $\3$, $\s$, $\spr$ and $\sppr$ with
$\3 \otimes \3 = \3_s \oplus \3_a \oplus \s \oplus \spr \oplus
\sppr,\quad {\rm and}\quad \spr \otimes \spr = \sppr$.
Under the group element corresponding to $c (a)$, $\spr \to \omega
(\omega^2) \spr$ and $\sppr \to \omega^2 (\omega) \sppr$.

Let $(x_1,x_2,x_3)$ and $(y_1,y_2,y_3)$ denote the basis vectors for
two $\3$'s. Then
\begin{eqnarray}
(\3 \otimes \3)_{\3s,a}  & = & ( x_2 y_3 \pm x_3 y_2\, ,\, x_3 y_1 \pm
x_1 y_3\, ,\nonumber\\
& \, & x_1 y_2 \pm x_2 y_1 ) \nonumber \\
(\3 \otimes \3)_{\s} & = & x_1 y_1 + x_2 y_2 + x_3 y_3 \nonumber\\
(\3 \otimes \3)_{\spr} & = & x_1 y_1 + \omega\, x_2 y_2 + \omega^2\, x_3 y_3 \nonumber\\
(\3 \otimes \3)_{\sppr} & = & x_1 y_1 + \omega^2\, x_2 y_2 +
\omega\, x_3 y_3
\end{eqnarray}

Under $SU(3) \otimes SU(2) \otimes U(1) \otimes A_4$, choose:\cite{xg}
\begin{eqnarray}
& Q_L \sim \left( 3,2,\frac{1}{3} \right) \left( \3 \right) & \nonumber\\
& u_R  \sim \left( 3,1,\frac{4}{3} \right)\left(\s \oplus \spr \oplus \sppr \right) &
\nonumber\\
&d_R  \sim \left( 3,1,-\frac{2}{3}
\right)\left(\s \oplus \spr \oplus \sppr \right)& \nonumber\\
& \ell_L \sim \left( 1,2,-1 \right) \left( \3 \right), & \nonumber\\ 
& \nu_R \sim \left( 1,1,0 \right)\left( \3 \right),& \nonumber\\
& e_R \sim \left( 1,1,-2 \right)\left(\s
\oplus \spr \oplus \sppr \right)&
\end{eqnarray}
for the fermions, and for the Higgs multiplets:
\begin{eqnarray}
& \Phi \sim \left( 1,2,-1 \right) \left( \3 \right),\ \phi \sim
\left( 1,2,-1 \right) \left( \s \right),& \nonumber \\
& \chi \sim \left(1,1,0 \right) \left( \3 \right).& 
\end{eqnarray}

The required spontaneous symmetry breaking pattern is given by
the VEVs:
\begin{eqnarray}
& \langle\Phi^0\rangle = (v,v,v),\qquad A_4 \to Z_3 & \nonumber\\
& \langle\chi\rangle = (0,v_\chi,0),\qquad A_4 \to Z_2 &\nonumber \\
& \langle\phi\rangle = v_\phi,\qquad A_4 \to A_4 &
\end{eqnarray}
The quark mass matrices come from $\langle\Phi\rangle$ and have the form
$U(\omega)$ multiplied by a diagonal matrix of arbitrary eigenvalues, so at tree level
$U_{CKM} = 1$. The
charged lepton mass matrices also come from $\langle\Phi\rangle$, so the left 
diagonalisation matrix is $U(\omega)$. The
neutrino Dirac masses arise from $\langle\phi\rangle$:  $m_\nu^D {\rm diag}(1,1,1)$.
The neutrino RH Majorana masses 
are driven by $\langle\chi\rangle$ plus bare masses. The overall $\nu$ mass matrix is
\begin{equation} 
\left( \begin{array}{cccccc}
0 & 0 & 0 & m_{\nu}^D & 0 & 0 \\
0 & 0 & 0 & 0 & m_{\nu}^D & 0 \\
0 & 0 & 0 & 0 & 0 & m_{\nu}^D \\
m_{\nu}^D & 0 & 0 & M & 0 & M_\chi \\
0 & m_{\nu}^D & 0 & 0 & M & 0 \\
0 & 0 & m_{\nu}^D & M_\chi & 0 & M
\end{array} \right),
\end{equation}
and the effective light $\nu$ mass matrix is
\begin{eqnarray}
& M_L & = - M_\nu^D M_R^{-1} (M_\nu^D)^T \\
& = & - \frac{(m_\nu^D)^2}{M}
\left( \begin{array}{ccc}
\frac{M^2}{M^2-M^2_\chi}  & 0 & - \frac{M M_\chi}{M^2-M^2_{\chi}} \\
0 & 1 & 0 \\
- \frac{M M_\chi}{M^2-M^2_{\chi}} & 0 & \frac{M^2}{M^2-M^2_\chi}
\end{array} \right).\nonumber
\end{eqnarray}
Note the $Z_2$ structure in $(1,3)$ subspace.

So, at tree-level we have tribimaimxal mixing (up to phases):
\begin{equation}
V_{MNSP} = U(\omega)^{\dagger} V_L^{\nu}
 = \left( \begin{array}{ccc}
\frac{2}{\sqrt{6}} & \frac{1}{\sqrt{3}} & 0 \\
-\frac{\omega^2}{\sqrt{6}} & \frac{\omega^2}{\sqrt{3}} & -\frac{e^{-i\pi/6}}{\sqrt{2}} \\
-\frac{\omega}{\sqrt{6}} & \frac{\omega}{\sqrt{3}} &
\frac{e^{i\pi/6}}{\sqrt{2}}
\end{array} \right)
\end{equation}
In the neutrino sector, the mixing pattern is driven by $\langle\chi\rangle: A_4 \to Z_2$.
For the rest of the fermions, the patterns are driven by $\langle\Phi\rangle: A_4 \to Z_3$.
This dual symmetry breaking structure gives trivial CKM and tribimaximal MNSP.
For theory as a whole, of course, $A_4 \to$ nothing. We can describe this
situation as ``parallel worlds of $A_4$ symmetry breaking.''\cite{xg}

\section{Corrections after flavour symmetry breaking}

The above mixing matrix results hold only at lowest order.  After spontaneous $A_4$
symmetry breaking, deviations are induced.  Because of
the parallel worlds of symmetry breaking, it is useful to classify these effects
into those within each sector (the neutrino sector and the charged-fermion sector),
and those acting between sectors.\cite{xg}  

Within each sector, we can write down the mass entries permitted by the unbroken symmetry
in that sector, not all of which are generated at tree-level.
For quarks and charged leptons, the tree-level form is not changed, so the 
left diagonalisation matrices are still $U(\omega)$.
This means the CKM matrix is still trivial.  This is ensured by the unbroken $Z_3$ in the quark sector.

But, the effective light $\nu$ matrix changes:
\begin{eqnarray}
M_L & \to & M_L \nonumber\\ 
& + & \left. \left( \begin{array}{ccc} \delta_{11} & 0 &
\delta_{13} \\ 0 & \delta_{22} & 0 \\ \delta_{13} & 0 &
\delta_{33}
\end{array} \right) \right|_{\rm h.o.}
\end{eqnarray}
where h.o.\ denotes higher order. This means that $V_L^{\nu}$ becomes
\begin{eqnarray}
\left( \begin{array}{ccc} 1 & 0 & 0 \\ 0 & 1 & 0 \\ 0
& 0 & e^{i\beta}
\end{array} \right) & \times &\nonumber\\
\left( \begin{array}{ccc} \cos\theta & \ 0 & \ -\sin\theta \\ 0 &
\ 1 & \ 0 \\ \sin\theta & \ 0 & \ \cos\theta
\end{array} \right) &\times& \nonumber\\
\left( \begin{array}{ccc} e^{i\alpha_1} & 0 & 0 \\ 0 &
e^{i\alpha_2} & 0 \\ 0 & 0 & e^{i\alpha_3}
\end{array} \right)& &
\end{eqnarray}
where $\theta = \frac{\pi}{4} + \delta$ and $|\delta| \ll 1$ if the h.o.\ corrections
are small.  Hence,
\begin{eqnarray}
& V_{MNSP} & = U(\omega)^{\dagger} V_L^{\nu} \\
& = & \frac{1}{\sqrt{3}}
\left( \begin{array}{ccc}
c + s e^{i\beta} & 1 & c e^{i\beta} - s \\
c + \omega s e^{i\beta} & \omega^2 & \omega c e^{i\beta} - s \\
c + \omega^2 s e^{i\beta} & \omega & \omega^2 c e^{i \beta} - s
\end{array} \right) \nonumber
\end{eqnarray}
There are deviations from tribimaximal mixing in the first and third columns, including
a nonzero $U_{e3}$.

We now turn to interactions between the sectors.
To generate realistic CKM mixing, we need to break the $Z_3$.  But in the theory overall,
it {\it is} broken.  Hence one way that CKM mixing might be generated is through the
mediation of $Z_3$ breaking in the neutrino sector to the quark sector, for
example through effective operators like
\begin{eqnarray}
& \overline{Q}_L \, u_R \, \Phi \, \chi,\ \overline{Q}_L \,
u'_R \, \Phi \, \chi,\ \overline{Q}_L \, u''_R \, \Phi \, \chi &
\nonumber\\
& \overline{Q}_L \, d_R \, \tilde{\Phi} \, \chi,\
\overline{Q}_L \, d'_R \, \tilde{\Phi} \, \chi,\
\overline{Q}_L \, d''_R \, \tilde{\Phi} \, \chi, &
\end{eqnarray}
There is enough freedom at this level to generate a realistic CKM matrix.
But it is not yet clear if that is the best way to do it, though it seems
like a natural feature to have.

\section{Challenges and conclusions}

The theory needs to be ``completed'', as the above is a symmetry scheme without
a fully sepcified dynamics. The default possibility is a standard Higgs
potential.  But this raises a non-trivial problem:
How to keep the parallel worlds of symmetry breaking controllably intact?
Unfortunately, the Higgs potential interactions between $\Phi$ and $\chi$ tend to spoil the
different VEV patterns required.
So, at least some of these interactions need to eliminated.  What are the logical possibilities?\cite{xg}

Normal internal symmetries do not work, because a term like $\Phi^{\dagger} \Phi \chi^2$
is always invariant.

One possibility is to decouple $\chi$ or $\nu_R, \chi$ from rest of theory by making
certain parameters very small (hidden sector).  It is not known if this can be made 
to work

A second possibility is supersymmetry, because terms like $\Phi^{\dagger} \Phi \chi^2$ come from
superpotential terms $\Phi_u\, \Phi_d\, \chi$ and the latter {\it can} be forbidden
by an internal symmetry.  We have constructed an inelegant existence proof for this.

A third possibility is to sequester $\chi$ on different brane from $\Phi$.\cite{af}

There may be others.

Overall, we conclude that $A_4$ has the potential to simultaneously explain the quark and
lepton mixing matrices, while leaving the masses arbitrary.  This works at a
symmetry level -- which is my main point -- but a dynamically complete theory
is a work-in-progress.

\section*{Acknowledgments}
This work was supported by the Australian Research Council.

\end{document}